\documentclass[a4paper,
               keeplastbox,   
               ]{jacow}
\usepackage{pdfpages,multirow,ragged2e} %
\makeatletter%
	\ifboolexpr{bool{xetex}}
	 {\renewcommand{\Gin@extensions}{.pdf,%
	                    .png,.jpg,.bmp,.pict,.tif,.psd,.mac,.sga,.tga,.gif,%
	                    .eps,.ps,%
	                    }}{}
\makeatother

\ifboolexpr{bool{xetex} or bool{luatex}} 
 {}                                      
 {\usepackage[utf8]{inputenc}}           

\usepackage[USenglish]{babel}

\ifboolexpr{bool{jacowbiblatex}}%
 {%
  \addbibresource{jacow-test.bib}
  \addbibresource{biblatex-examples.bib}
 }{}
\graphicspath{{./image/}}

\begin{document}

\title{Design of a HOM-Damped 166.6 MHz Compact Quarter-Wave Beta=1 Superconducting Cavity for High Energy Photon Source}

\author{X.Y.~Zhang\textsuperscript{1}\thanks{Email: zhangxinying@ihep.ac.cn}, J.~Dai\textsuperscript{1}, L.~Guo\textsuperscript{1,2}, T.M.~Huang\textsuperscript{1,2}, Z.Q.~Li\textsuperscript{1,2}, Q.~Ma\textsuperscript{1,2},\\  F.B.~Meng\textsuperscript{1}, Z.H.~Mi\textsuperscript{1,2}, P.~Zhang\textsuperscript{1,2}, H.J.~Zheng\textsuperscript{1} \\
\textsuperscript{1}Institute of High Energy Physics, Chinese Academy of Sciences, Beijing 100049, China \\
\textsuperscript{2}University of Chinese Academy of Sciences, Beijing 100049, China}
	
\maketitle

\begin{abstract}
Superconducting cavities with low RF frequencies and heavy damping of higher order modes (HOM) are desired for the main accelerator of High Energy Photon Source (HEPS), a 6~GeV synchrotron light source promising ultralow emittance currently under construction in Beijing. A compact 166.6~MHz superconducting cavity was proposed adopting a quarter-wave beta=1 geometry. Based on the successful development of a proof-of-principle cavity, a HOM-damped 166.6~MHz compact superconducting cavity was subsequently designed. Ferrite damper was installed on the beam pipe to reduce HOM impedance below stringent threshold of coupled-bunch instabilities. Being compact, RF field heating on the cavity vacuum seal was carefully examined against quenching the NbTi flange. The cavity was later dressed with helium vessel and the tuning mechanism was also realized. Excellent RF and mechanical properties were eventually achieved. Finally, the two-cavity string was designed to ensure smooth transitions among components and proper shielding of synchrotron light. This paper presents a complete design of a fully dressed HOM-damped low-frequency beta=1 superconducting cavity for HEPS.
\end{abstract}

\section{Introduction}
High Energy Photon Source (HEPS) is 6 GeV diffraction limited synchrotron light source under construction in Beijing\cite{Y-Jiao-IPAC21}. Its main parameters of HEPS storage ring are listed in Table.~\ref{tab:para}. A modified hybrid seven-bend achromat~(7BA) lattice has been designed to push the natural beam emittance down to 34 pm while preserving a high brightness of the X-ray synchrotron light. For the storage ring with a circumference of 1360.4 m, five 166.6-MHz superconducting cavities have been adopted as fundamental rf accompanied by two 499.8-MHz superconducting cavities as third harmonic. This configuration is to accommodate a novel injection scheme for the future while considering the technology readiness of both the rf cavity and the fast kicker\cite{Zhang:IPAC21-MOPAB380}. The choice of frequency was extensively elucidated elsewhere from both physics\cite{Y-Jiao-IPAC21} and technology aspects\cite{PoP-1}. The fundamental cavity makes use of a superconducting quarter-wave beta=1 structure due to a low operating frequency.

\begin{table}[!hbt]
\centering
\caption{Main parameters of the HEPS storage ring.}
\begin{tabular}{lcc}
\toprule
\textbf{Parameter} & \textbf{Value} & \textbf{Unit} \\
\midrule
Circumference & 1360.4 & m   \\ 
Beam energy & 6 & GeV   \\ 
Beam current & 200 & mA \\
Energy loss per turn (w/IDs) & 4.4 & MV \\
Total beam power & 900 & kW \\
Fundamental RF frequency & 166.6 & MHz \\
Total RF voltage & 5.4 & MV \\
3$^\text {rd}$ harmonic rf frequency & 499.8 & MHz \\
\bottomrule
\end{tabular}
\label{tab:para}
\end{table}

A proof-of-principle (PoP) cavity has been previously developed between 2016 and 2019 and its performances have been extensively studied in both vertical tests and horizontal tests\cite{PoP-1,PoP-2,PoP-3,PoP-4}. The development of the PoP cavity explored the design of the cavity's main body, verified the feasibility of manufacturing and related processes, transformed and developed a post-processing system, and analyzed the test results. This provided a lot of experience and lessons for us. Based on the successful development of a proof-of-principle cavity, a HOM-damped 166.6 MHz compact superconducting cavity was subsequently designed.

The electromagnetic (EM) design, multipacting (MP), mechanical design, fabrication, and layouts of cavity string of the HOM-damped cavity are described in the following sections.

\section{The cavity design}
The specifications of HEPS 166.6~MHz HOM-damped superconducting cavities are listed in Table.~\ref{tab:spe}. The design will focus on the HOM damped design, meanwhile, the successful part of the PoP cavity should be kept as much as possible, and the problems found in the test should be improved.

\begin{table}[!hbt]
\centering
\caption{Cavity specifications.}
\begin{tabular}{lcc}
\toprule
\textbf{Parameter} & \textbf{Value} & \textbf{Unit} \\
\midrule
Frequency & 166.6 & MHz   \\ 
Design voltage $V_d$             & 1.5               & MV\\
$Q_0$ at $V_d$                   & \textgreater 1$\times$10$^9$   &- \\
$E_{peak}$ at $V_d$              & \textless40              & MV/m\\
$B_{peak}$ at $V_d$              & \textless65              & mT\\
Operating mode                    & CW      &-\\
Loaded Q                    & 5$\times$10$^4$       &-\\
Loaded bandwidth                   & 3.33      &kHz\\
Operating temperature         & 4.2 & K   \\ 
Operating LHe pressure                  & 1230±3 & mbar   \\ 
Sensitivity to LHe             & \multirow{2}{*}{\textless27.5}  & \multirow{2}{*}{Hz/mbar}  \\ 
pressure($df/dp$)&&\\
LFD coeff. &\textless5&Hz/(MV/m)$^2$\\
Longitudinal stiffness &\textless10&kN/mm\\
Frequency of lowest&\multirow{2}{*}{\textgreater 100} &\multirow{2}{*}{ Hz}\\
mechanical mode&&\\
Buckling critical factor&\multirow{2}{*}{\textgreater2.0/16.1} &\multirow{2}{*}{ -}\\
(cylinder/cone)&&\\
Remnant H field&\textless4&$\mu$T\\
\bottomrule
\end{tabular}
\label{tab:spe}
\end{table}

\subsection{The EM design}
\begin{figure}[!htb]
\centering
\includegraphics*[width=\columnwidth]{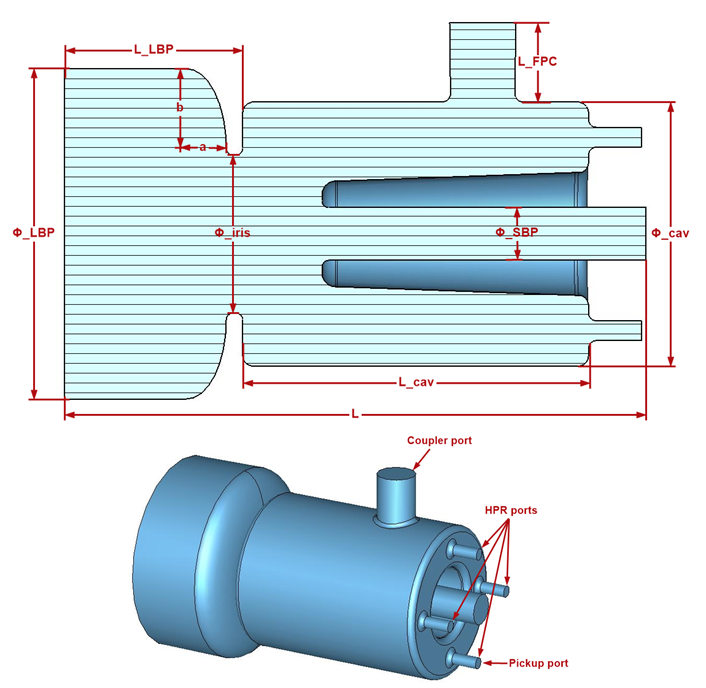}
\caption{The HOM-damped 166.6 MHz cavity.}
\label{fig:CAV}
\end{figure}

Since the cavity frequency is only 166.6 MHz, the most compact quarter-wave cavity structure is uesed as the main body of the cavity to reduce the size to the processable. The success of the PoP cavity test indicates that the cavity structure has excellent rf performance. The optimized cavity geometry is shown in Fig.~\ref{fig:CAV}, and the parameter differences with the PoP cavity are listed in Table.~\ref{tab:geo}. 

\begin{table*}[!hbt]
\centering
\caption{Comparison of cavity geometry parameters. Fund. mode stands for fundamental mode.}
\begin{tabular}{lcccc}
\toprule
\textbf{Parameter} & \textbf{Unit} & \textbf{PoP Cav.} & \textbf{HOM-damped Cav.} & \textbf{remarks}\\
\midrule
Length of flange to flange(L) & mm&853 & 880 &\multirow{3}{*}{Reduce power loss of rf leakage field}    \\
Length of LBP(L$ \text{\_LBP}$) & mm&250 & 270 &\\ 
Length of FPC(L$ \text{\_FPC}$) & mm&80 & 120 &  \\ 
Length of cavity(L$ \text{\_cav}$)& mm&   \multicolumn{2}{c}{523} &-  \\ 
Cavity diameter($\Phi \text{\_cav}$) & mm&390 & 400 &Optimize EM parameters  \\  
Aperture diameter($\Phi \text{\_LBP}$) & mm&170 & 505 & \multirow{2}{*}{Propagate HOMs/reject Fund. mode}\\
Aperture diameter($\Phi \text{\_iris}$) & mm&170 & 240&  \\ 
Aperture diameter($\Phi \text{\_SBP}$) & mm&   \multicolumn{2}{c}{80} &-  \\ 
Transition of the LBP(a, b) & mm&- & 60, 120 &Easy to tune \\ 
\bottomrule
\end{tabular}
\label{tab:geo}
\end{table*}

The main body of the cavity was basically the same as that of the PoP. Only the diameter of the cavity ($\Phi \text{\_cav}$) has been slightly increased to 400mm to further improve EM performance.

Based on the PoP cavity, the cavity's structure was further optimized to solve the HOM problem. The aperture diameter of the cavity left beam pipe ($\Phi \text{\_LBP}$) and the aperture diameter of the cavity iris ($\Phi \text{\_iris}$) were increased from 170 mm to 505 mm and 240 mm, respectively, to propagate the HOMs field and reject the fundamental mode. Ferrite absorber will be installed on the enlarged beam pipe extension at room temperature to extract the HOM power. The whole cavity impedance spectrum was simulated by CST Studio Suite\cite{CST} and ABCI software\cite{ABCI}. All the impedance of HOMs are under the impedance threshold, except for the second monopole (M2) mode. The impedance for M2 mode is 1.28E4 $\Omega$ while the impedance threshold is 1.14E4 $\Omega$. A detailed introduction of these impedance results and power losses due to HOMs are given in \cite{HOM}.

\begin{figure}[!htb]
\centering
\includegraphics*[width=0.8\columnwidth]{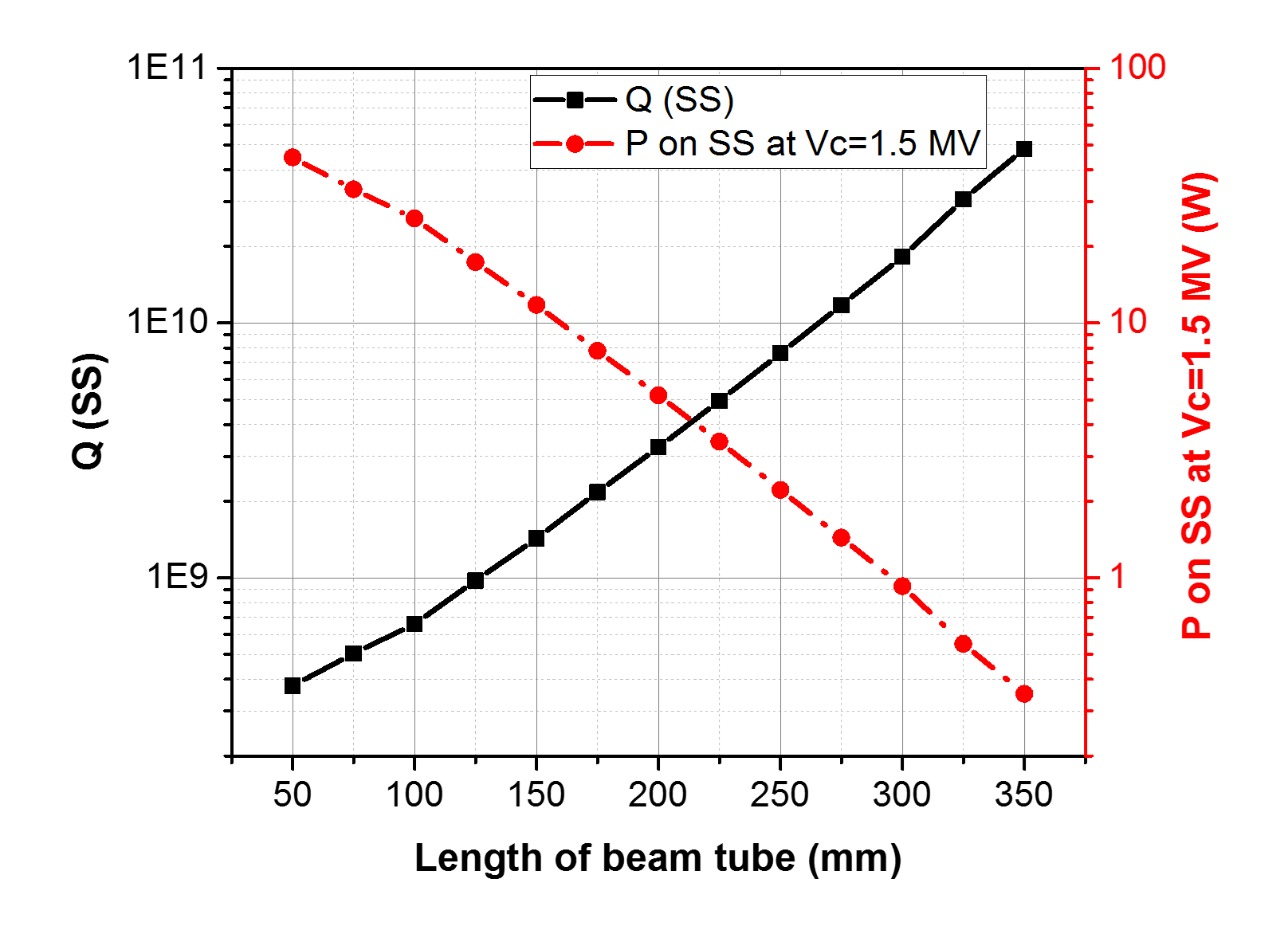}
\caption{Q and power loss on stainless steel transition tube.}
\label{fig:Ploss}
\end{figure}

After the beam pipe diameter of the cavity is increased, the length of the pipe also needs to be reconfirmed. The stainless steel (SS) transition tube is connected to the niobium tube of the cavity. In order to fully attenuate the fundamental electromagnetic field of the cavity and reduce the power dissipation of the leakage field on the SS transition tube, the large beam pipe of the superconducting cavity should be long enough. Otherwise this will reduce the $Q_0$ of the cavity operating mode, heat up the cavity peripherals, and introduce a significant additional load of the system. The simulation results are shown in Fig.~\ref{fig:Ploss}. When the $Q_0$ on the stainless steel transition tube reaches 1e10, which is an order of magnitude higher than the cavity design 1e9, it is considered that the superconducting cavity rf performance will not be affected. Due to the longitudinal compact design requirements of the cavity string, the length of the large beam pipe is chosen to be 270 mm.

An overheating at the cavity-coupler interface region was observed in high-power tested, causing a “thermal runaway” and eventually quenched the cavity. Solutions were proposed including an elongated niobium extension tube from 80 mm to 120 mm at the coupler port and an optimized helium gas cooling of the power coupler’s outer conductor. More details can be found in \cite{PoP-4}.

The bend of the LBP transition has been changed from R20 mm to an elliptical arc (60 mm, 120 mm)  for the purpose of enhance the tuning ability.

Due to the enlarged cavity beam pipe and a accommodated large helium jacket, the original electric-probe pickup of PoP becomes inevitably long with unfavorable mechanical properties. Relocated to an existing high-pressure-rinsing (HPR) port, a low-loss magnetic-loop pickup was designed. The consequent removal of the original pickup port from the cavity largely simplified the helium jacket fabrication and cavity contamination\cite{166SCC-pickup}. 

\begin{figure}[!htb]
\centering
\includegraphics*[width=0.8\columnwidth]{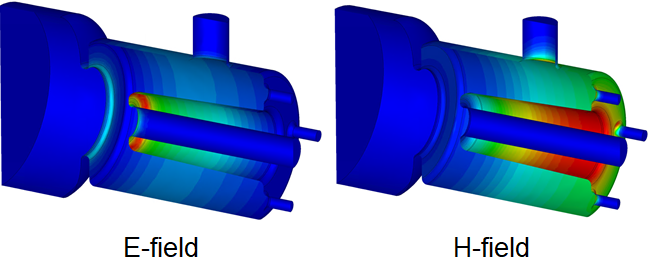}
\caption{The distribution of electromagnetic field.}
\label{fig:EM-field}
\end{figure}

The electromagnetic field distribution of the fundamental mode is shown in Fig.~\ref{fig:EM-field} and the RF parameters are listed in Table.~\ref{tab:rf}. A 20\% margin on peak surface fields was reserved.

\begin{table}[!hbt]
\centering
\caption{The RF parameters.}
\begin{tabular}{lcc}
\toprule
\textbf{Parameter} & \textbf{Value} & \textbf{Unit} \\
\midrule
$R/Q$ (=$|V|^2$/$\omega$U)                 & 139             & $\Omega$ \\ 
$G$ (=$R_s\cdot Q_0$) & 56     & $\Omega$ \\
$E_{peak}$ at $V_d$=1.5MV              & 40              & MV/m\\
$B_{peak}$ at $V_d$=1.5MV              & 62              & mT\\
\bottomrule
\end{tabular}
\label{tab:rf}
\end{table}

\subsection{Multipacting}
\begin{figure}[!htb]
\centering
\includegraphics*[width=.8\columnwidth]{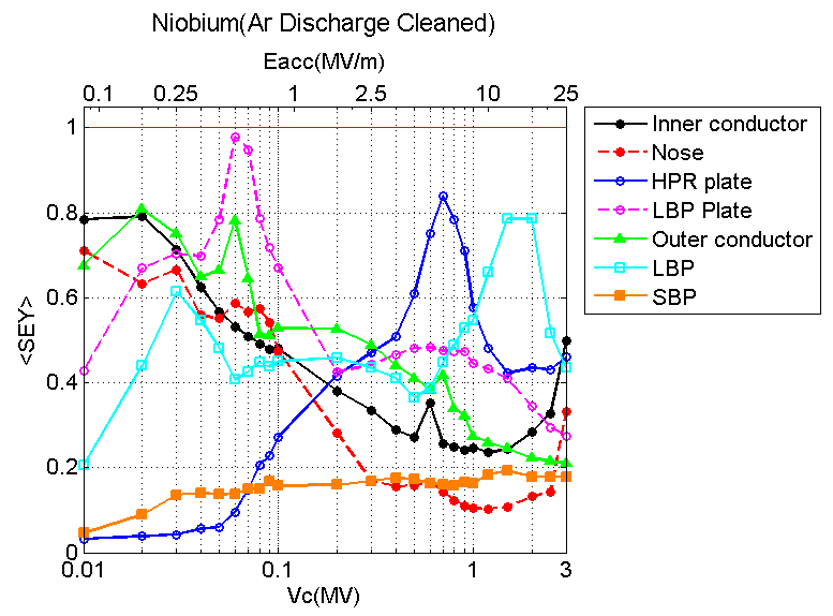}
\caption{<SEY> curves for different regions of the cavity.}
\label{fig:MP}
\end{figure}

CST Particle Studio and CST Microwave Studio\cite{CST} were used for multipacting simulation of the 166 MHz HOM-damped SRF cavity. Argon discharge cleaned niobium was selected, which has a low secondary emission coefficient. In order to speed up the computation and to determine MP locations, the cavity inner surface is divided into several initial particle source regions. Particles are randomly emitted in all directions with the initial energy uniformly distributed from 0 to 4 eV. Particle number, trajectories and energies are tracked over 10 rf periods in fields with different acceleration voltages and relative phases. The normalized secondary electron yield (<SEY>) is defined to evaluate the probability of multipacting. Fig.~\ref{fig:MP} shows the <SEY> curves at different regions and voltages. Two soft multipacting bands at low field (0.06 MV) on the LBP plate and at medium field (0.6 MV) on the HPR plate are similar to those of the PoP cavity. A new soft band near the design voltage of 1.5 MV appears at the LBP transition region. The <SEY> are all less than 1, indicating that multipacting doesn’t occur when the cavity inner surface is well treated.

\subsection{The mechanical design}
\begin{figure}[!htb]
\centering
\includegraphics*[width=.8\columnwidth]{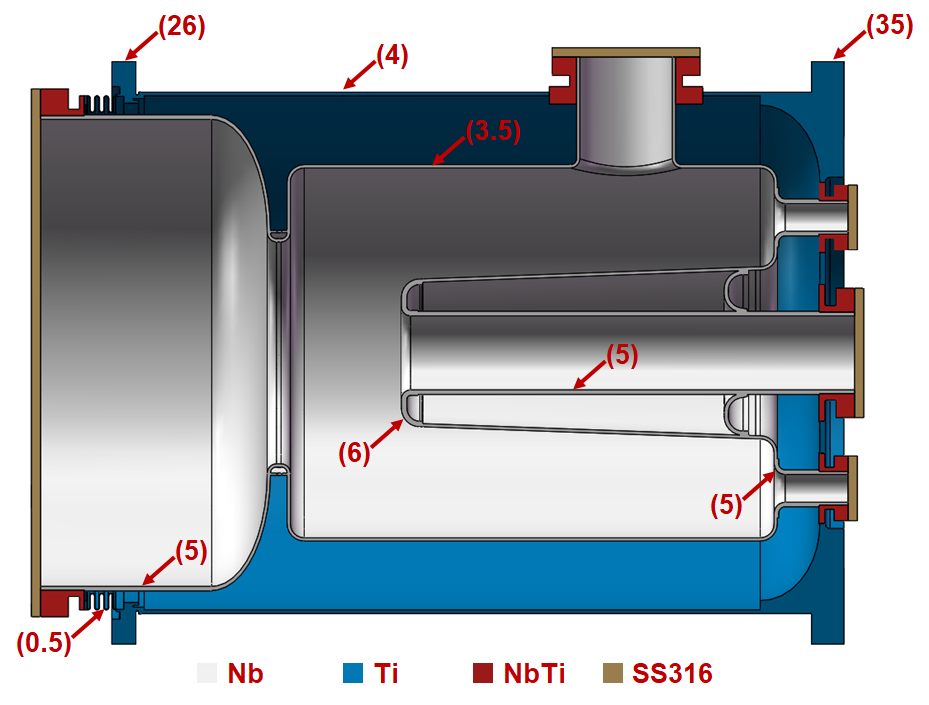}
\caption{The thickness and material of the dressed cavity.}
\label{fig:Thi}
\end{figure}

The mechanical design optimization of the 166.6~MHz dressed cavity was conducted by using ANSYS software\cite{ANSYS} and CST Studio Suite\cite{CST}. The model of the dressed cavity is shown in Fig.~\ref{fig:Thi}. The wall thickness of cavity varies from 3.5 mm to 6 mm for different components. The thickness of Liquid Helium (LHe) vessel, bellow and end plates were chosen to be 4 mm, 0.5 mm, 26 mm and 35 mm. The dressed cavity will be made of high-purity niobium (Nb) with RRR>300, the cavity flanges with Titanium Niobium alloy (NbTi), the stiffener with reactor grade niobium, and finally the LHe vessel and bellow with Grade-2 Titanium (Ti). The properties of these materials are exerted from Fermilab specification document: 5500.000-ES-371110. The maximum allowable stress value was established. ASME Boiler and Pressure Vessel Code and Chinese code were also followed. A detailed discussion can be found in a previous publication\cite{PoP-2}. Compared with the PoP cavity, a large helium jacket was required to accommodate the HOM-damped cavity with the enlarged LBP. In order to reduce the cavity stiffness to an acceptable range, the bellows was an unavoidable choice, although it will cause an increase in the cavity's peak stress and sensitivity to LHe pressure ($df/dp$). The bend and thickness of the niobium cavity HPR, the bend of junctions between the end plates and cylinder of the LHe vessel, and the structure and position of the LHe bellows were optimized. 

\begin{figure}[!htb]
\centering
\includegraphics*[width=.8\columnwidth]{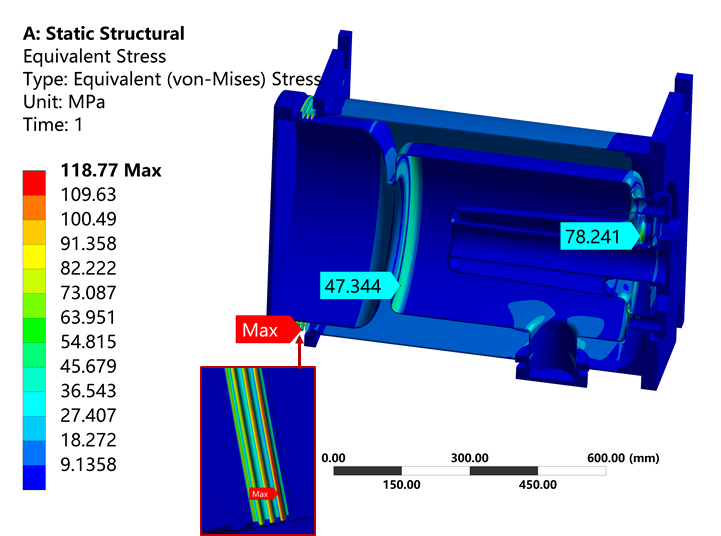}
\caption{The stress distribution of the entire structure under 1.5 atm pressure.}
\label{fig:Str}
\end{figure}

The optimized results of mechanical properties are summarized in Table.~\ref{tab:mechanical-results}. All the results were simulated under supports fixed. The stress distribution of the entire structure under 1.5 atm pressure inside the helium jacket volume is shown in Fig.~\ref{fig:Str}. The maximum stress reads 78~MPa on cavity  marginally exceeding the required value of 70.5~MPa at 293~K and 119~MPa on LHe bellow fulfilling the allowable value. The frequency sensitivity and longitudinal stiffness have eventually been reduced to -18~Hz/mbar and 9.7~kN/mm, respectively. The Lorentz force detuning coefficient is -1.1~Hz/(MV/m)$^2$, similar to PoP cavity. Meanwhile, the frequency of lowest mechanical mode and buckling critical factor also meet the design requirements.

\begin{table}[!hbt]
\centering
\caption{The mechanical parameters.}
\begin{tabular}{lcc}
\toprule
\textbf{Parameter} & \textbf{Value} & \textbf{Unit} \\
\midrule
Max. stress on cavity                 & 78            & MPa \\ 
Max. stress on bellow                 & 119            & MPa \\ 
$df/dp$                                 & -18            & Hz/mbar \\ 
$\Delta f$ under ±3mbar                   & ±55            & Hz \\ 
Longitudinal stiffness          & 9.7&kN/mm\\
Tuning range at 4.2 K         & 67&Hz\\
LFD coeff. &-1.1&Hz/(MV/m)$^2$\\
Frequency of lowest&\multirow{2}{*}{110} &\multirow{2}{*}{ Hz}\\
mechanical mode&&\\
Buckling critical factor&\multirow{2}{*}{11.4/73} &\multirow{2}{*}{ -}\\
(cylinder/cone)&&\\
\bottomrule
\end{tabular}
\label{tab:mechanical-results}
\end{table}

\begin{figure}[!htb]
\centering
\includegraphics*[width=.8\columnwidth]{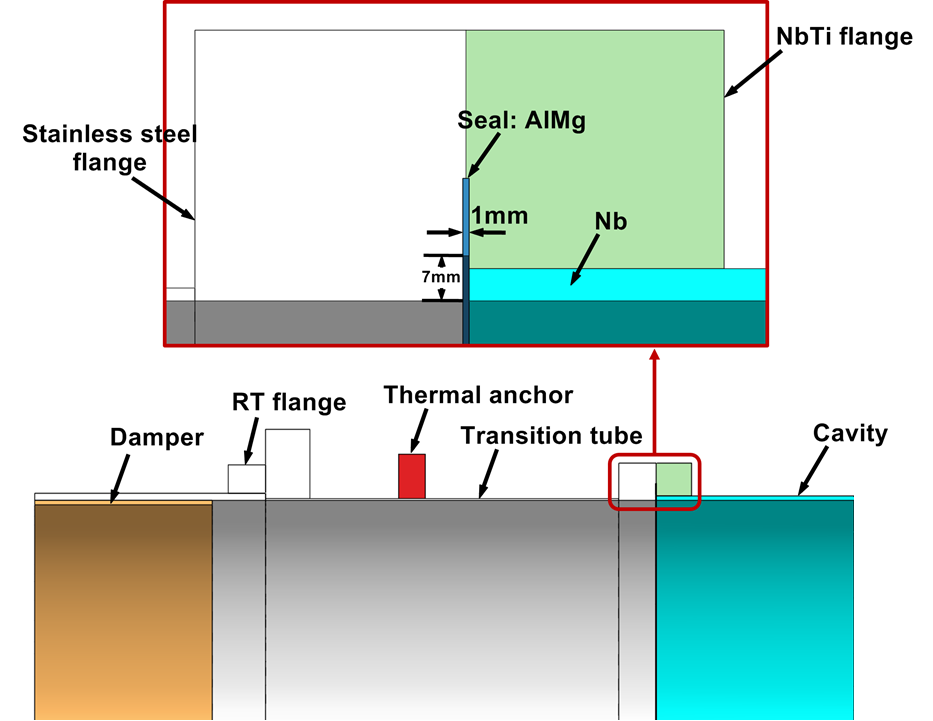}
\caption{Model of the sealing groove between cavity and transition tube.}
\label{fig:sealing-groove}
\end{figure}

The rf-thermal coupling simulation at the aluminum-magnesium(AlMg) alloy sealing groove was performed to confirm the temperature safety of the superconducting materials such as the NbTi flange and the niobium tube. The model of the sealing groove was shown in Fig.~\ref{fig:sealing-groove}. Peak temperatures of NbTi flange, niobium tube, AlMg sealing ring and SS flange are simulated to be 4.52 K, 4.32 K, 5.44 K and 6.63 K, respectively. Since the temperature of NbTi and Nb is less than the critical superconducting temperature, it will not cause quenching, so the length of LBP (270 mm) is feasible.

 The magnetic shielding has been designed for 166~MHz HOM-damped cavity to reduce the residual surface resistance due to magnetic flux trapping during cavity cool down. A thin layer (1.3 mm) of permalloy (1J79) was designed and composed of three parts inside the jacket. Final equivalent contribution for residual resistance caused by geomagnetic field and magnet fringe field was calculated to be 2.36 n$\Omega$, and the equivalent residual magnet field on cavity wall was 1.93~$\mu$T. More information was presented in detail in \cite{166SCC-Magnetic-Shielding}  

\section{The fabrication}
\begin{figure}[!htb]
\centering
\includegraphics*[width=\columnwidth]{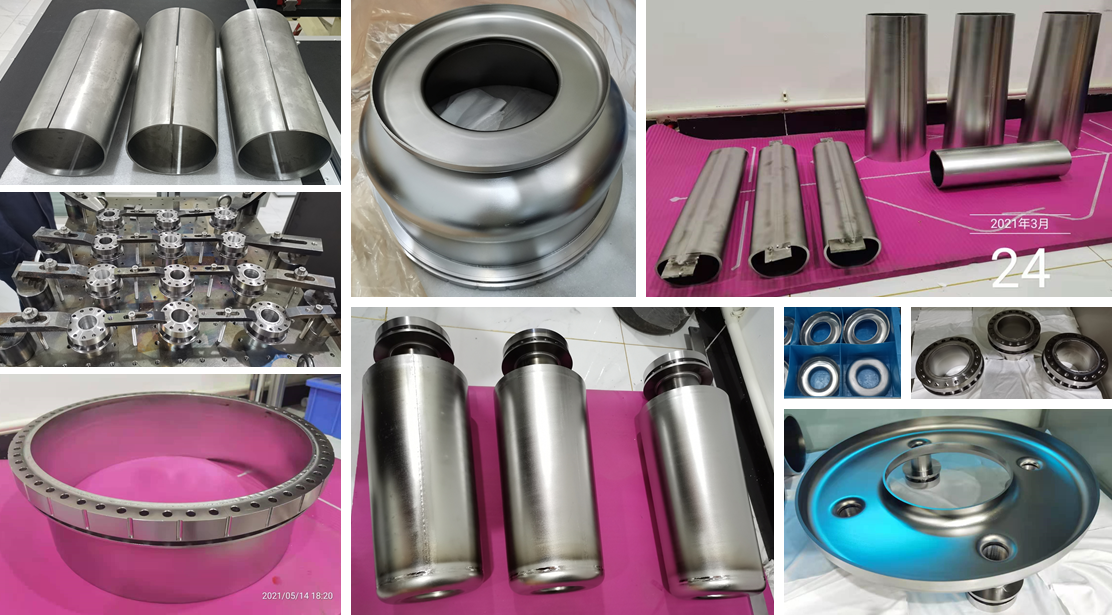}
\caption{ The fabrication of the HOM-damped cavity.}
\label{fig:Fab}
\end{figure}

The dressed cavity is contracted to Beijing HE-Racing Technology Co., Ltd. Some of the key fabrication processes and components are shown in Fig.~\ref{fig:Fab}. Deep drawing, rolling, machining, and electron-beam welding are the main manufacturing techniques used in the fabrication of this cavity. The prototype HOM-damped 166~MHz cavity will be manufactured in July 2021, and vertical test is planned for September.

\section{The cavity string}
\begin{figure}[!htb]
\centering
\includegraphics*[width=\columnwidth]{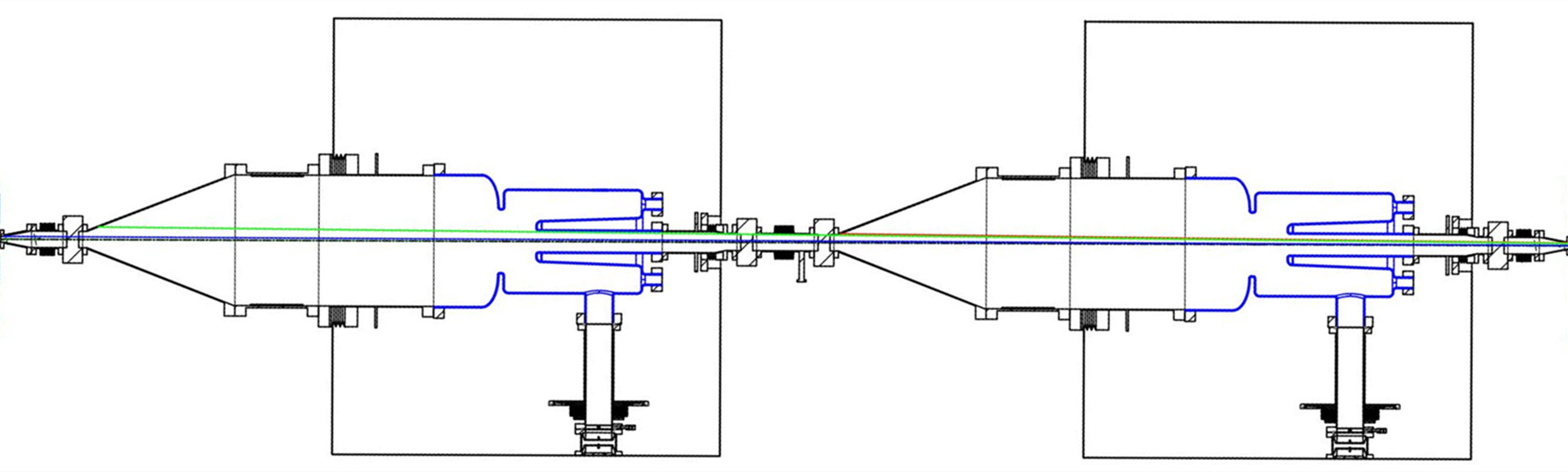}
\caption{ The layout of the cavity string.}
\label{fig:lay}
\end{figure}

A HOM-damped 166.6~MHz quarter-wave superconducting cavity string will be used for the HEPS storage ring and fit into a straight section of only 6 meters. The large size of the low-frequency cavity, the increased large beam pipe for the HOM, the transition tube, the smoother taper and synchrotron light shielding pose severe challenges to the limited straight section. Four layouts for the cavity string and the collimator design were presented in \cite{166SCC-Synchrotron-Light-Shielding}. Considering the replaceability of the cryomodule, the broadband impedance of the cavity string, synchronous light irradiation and future maintainability, the solution of placing two cryomodules in sequence was finally selected, in which the beam enters from the small beam pipe, as shown in Fig.~\ref{fig:lay}. The loss factor of the cavity string was 5.324 V/pC, and two collimators were design to successfully blocked the synchronous light.

\section{Final Remarks}
A HOM-damped 166.6~MHz compact quarter-wave beta=1 superconducting cavity has been proposed for High Energy Photon Source. Based on the PoP cavity, the cavity’s structure was further optimized, and the HOM impedances can be reduced below the threshold by a ferrite absorber mounted on the enlarged left beam pipe. Excellent rf and mechanical properties were eventually achieved. The layout of cavity string in rf section was also analyzed. The cavity is being manufactured and will be tested in the fall of 2021.

\section{ACKNOWLEDGEMENTS}
This work was supported in part by High Energy Photon Source, a major national science and technology infrastructure in China and in part by the National Natural Science Foundation of China under Grant 12005241.


%
%
\ifboolexpr{bool{jacowbiblatex}}%
	{\printbibliography}%

\begin{thebibliography}{9} 
	
	
	\bibitem{Y-Jiao-IPAC21}
   	Y. Jiao \emph{et al.}, \textquotedblleft{Progress of Lattice Design and Physics Studies on the High Energy Photon Source}\textquotedblright, presented at the 12th Int. Particle Accelerator Conf. (IPAC'21), Campinas, Brazil, May 2021, paper MOPAB053.            
	
	\bibitem{Zhang:IPAC21-MOPAB380}
   P. Zhang \emph{et al.}, \textquotedblleft{Status and Progress of the RF System for High Energy Photon Source}\textquotedblright,
   presented at the 12th Int. Particle Accelerator Conf. (IPAC’21), Campinas, Brazil, May 2021, paper MOPAB380.     

	\bibitem{PoP-1}
		P.~Zhang, \textit{et al.}, \textquotedblleft{Development and vertical tests of a 166.6 MHz proof-of-principle superconducting quarter-wave beta = 1 cavity}\textquotedblright, \textit{Rev. Sci. Instrum.} 90, 084705 (2019). \url{https://doi.org/10.1063/1.5119093}

       	\bibitem{PoP-2}
		X. Y.~Zhang, \textit{et al.}, \textquotedblleft{The mechanical design, fabrication and tests of a 166.6 MHz quarter-wave beta=1 proof-of-principle superconducting cavity for HEPS}\textquotedblright, \textit{NIM-A} 947, 162770 (2019). \url{https://doi.org/10.1016/j.nima.2019.162770}

	\bibitem{PoP-3}
		X. Y.~Zhang, \textit{et al.}, \textquotedblleft{Design and Mechanical Performance of a Dressed 166.6 MHz $\beta$=1 Proof-of-Principle Superconducting Cavity in Horizontal Tests}\textquotedblright, \textit{IEEE Trans. Appl. Supercond.} 30(8), 3500208 (2020). \url{https://doi.org/10.1109/TASC.2020.2999541}
     
       \bibitem{PoP-4}
		T. M.~Huang, \textit{et al.}, \textquotedblleft{High-power tests and solution to overheating at cavity-coupler interface of the 166.6-MHz beta=1 superconducting quarter-wave resonator for HEPS}\textquotedblright, \textit{AIP Advances} 11, 045024 (2021). \url{https://doi.org/10.1063/5.0046377}

        \bibitem{CST}
                 CST Studio Suite\textregistered, Version 2020, CST AG, Darmstadt, Germany.
                 
        \bibitem{ABCI}
                 Y.H. Chin, ABCI, https://abci.kek.jp.

      \bibitem{HOM}
               	H. J.~Zheng \emph{et al.}, \textquotedblleft{Higher order mode damping for 166 MHz and 500 MHz superconducting RF cavities at High Energy Photon Source}\textquotedblright, presented at the 12th Int. Particle Accelerator Conf. (IPAC’21), Campinas, Brazil, May 2021, paper WEPAB090.            

        \bibitem{166SCC-pickup}
		      T.M.~Huang, \textit{et al.}, \textquotedblleft{Development of a low-loss magnetic-coupling pickup for 166.6 MHz quarter-wave beta=1 superconducting cavities}\textquotedblright, \textit{NUCL SCI TECH} 31, 87 (2020). \url{https://doi.org/10.1007/s41365-020-00795-6}

        \bibitem{ANSYS}
                 ANSYS (Release 17), http://www.ansys.com. 

	\bibitem{166SCC-Magnetic-Shielding}
   	        L.~Guo \emph{et al.}, \textquotedblleft{Design of the Magnetic Shielding for 166-MHz and 500-MHz Superconducting RF Cavities at High Energy Photon Source}\textquotedblright, presented at the 12th Int. Particle Accelerator Conf. (IPAC’21), Campinas, Brazil, May 2021, paper TUPAB340.        
     
      \bibitem{166SCC-Synchrotron-Light-Shielding}
                X. Y. Zhang, Z. Q. Li, Q. Ma, and P. Zhang, \textquotedblleft{Synchrotron Light Shielding for the 166 MHz Superconducting RF Section at High Energy Photon Source}\textquotedblright, presented at the 12th Int. Particle Accelerator Conf. (IPAC’21), Campinas, Brazil, May 2021, paper MOPAB382.            
	

	\end{thebibliography}
	{%
	
	
} 
%
%


\end{document}